\documentclass[useAMS,usenatbib]{mn2e}

\usepackage{defs}
\usepackage{psfig}
\usepackage{amsmath}
\usepackage{amssymb}

\title[HerMES: SPIRE/Sub-millimetre Emission from Radio Selected AGN]{HerMES: SPIRE Emission from Radio Selected AGN\thanks{Herschel is an ESA space observatory with science instruments provided by European-led Principal Investigator consortia and with important participation from NASA.}}

\author[N.~Seymour et al.]
{\parbox{\textwidth}{\raggedright N.~Seymour,$^{1}$\thanks{E-mail: \texttt{nps@@mssl.ucl.ac.uk}}
M.~Symeonidis,$^{1}$
M.J.~Page,$^{1}$
A.~Amblard,$^{2}$
V.~Arumugam,$^{3}$
H.~Aussel,$^{4}$
A.~Blain,$^{5}$
J.~Bock,$^{5,6}$
A.~Boselli,$^{7}$
V.~Buat,$^{7}$
N.~Castro-Rodr{\'\i}guez,$^{8,9}$
A.~Cava,$^{8,9}$
P.~Chanial,$^{4}$
D.L.~Clements,$^{10}$
A.~Conley,$^{11}$
L.~Conversi,$^{12}$
A.~Cooray,$^{2,5}$
C.D.~Dowell,$^{5,6}$
E.~Dwek,$^{13}$
S.~Eales,$^{14}$
D.~Elbaz,$^{4}$
A.~Franceschini,$^{15}$
J.~Glenn,$^{11}$
E.A.~Gonz\'alez~Solares,$^{16}$
M.~Griffin,$^{14}$
E.~Hatziminaoglou,$^{18}$
E.~Ibar,$^{19}$
K.~Isaak,$^{14}$
R.J.~Ivison,$^{19,3}$
G.~Lagache,$^{20}$
L.~Levenson,$^{5,6}$
N.~Lu,$^{5,21}$
S.~Madden,$^{4}$
B.~Maffei,$^{22}$
G.~Mainetti,$^{15}$
L.~Marchetti,$^{15}$
H.T.~Nguyen,$^{6,5}$
B.~O'Halloran,$^{10}$
S.J.~Oliver,$^{23}$
A.~Omont,$^{24}$
P.~Panuzzo,$^{4}$
A.~Papageorgiou,$^{14}$
C.P.~Pearson,$^{25,26}$
I.~P{\'e}rez-Fournon,$^{8,9}$
M.~Pohlen,$^{14}$
J.I.~Rawlings,$^{1}$
D.~Rizzo,$^{10}$
I.G.~Roseboom,$^{23}$
M.~Rowan-Robinson,$^{10}$
B.~Schulz,$^{5,21}$
Douglas~Scott,$^{17}$
D.L.~Shupe,$^{5,21}$
A.J.~Smith,$^{23}$
J.A.~Stevens,$^{27}$
M.~Trichas,$^{28}$
K.E.~Tugwell,$^{1}$
M.~Vaccari,$^{15}$
I.~Valtchanov,$^{12}$
L.~Vigroux,$^{24}$
L.~Wang,$^{23}$
G.~Wright,$^{19}$
C.K.~Xu$^{5,21}$ and
M.~Zemcov$^{5,6}$}\vspace{0.4cm}\\
\parbox{\textwidth}{\raggedright $^{1}$Mullard Space Science Laboratory, University College London, Holmbury St. Mary, Dorking, Surrey RH5 6NT, UK\\
$^{2}$Dept. of Physics \& Astronomy, University of California, Irvine, CA 92697, USA\\
$^{3}$Institute for Astronomy, University of Edinburgh, Royal Observatory, Blackford Hill, Edinburgh EH9 3HJ, UK\\
$^{4}$Laboratoire AIM-Paris-Saclay, CEA/DSM/Irfu - CNRS - Universit\'e Paris Diderot, CE-Saclay, pt courrier 131, F-91191 Gif-sur-Yvette, France\\
$^{5}$California Institute of Technology, 1200 E. California Blvd., Pasadena, CA 91125, USA\\
$^{6}$Jet Propulsion Laboratory, 4800 Oak Grove Drive, Pasadena, CA 91109, USA\\
$^{7}$Laboratoire d'Astrophysique de Marseille, OAMP, Universit\'e Aix-marseille, CNRS, 38 rue Fr\'ed\'eric Joliot-Curie, 13388 Marseille cedex 13, France\\
$^{8}$Instituto de Astrof{\'\i}sica de Canarias (IAC), E-38200 La Laguna, Tenerife, Spain\\
$^{9}$Departamento de Astrof{\'\i}sica, Universidad de La Laguna (ULL), E-38205 La Laguna, Tenerife, Spain\\
$^{10}$Astrophysics Group, Imperial College London, Blackett Laboratory, Prince Consort Road, London SW7 2AZ, UK\\
$^{11}$Dept. of Astrophysical and Planetary Sciences, CASA 389-UCB, University of Colorado, Boulder, CO 80309, USA\\
$^{12}$Herschel Science Centre, European Space Astronomy Centre, Villanueva de la Ca\~nada, 28691 Madrid, Spain\\
$^{13}$Observational  Cosmology Lab, Code 665, NASA Goddard Space Flight  Center, Greenbelt, MD 20771, USA\\
$^{14}$Cardiff School of Physics and Astronomy, Cardiff University, Queens Buildings, The Parade, Cardiff CF24 3AA, UK\\
$^{15}$Dipartimento di Astronomia, Universit\`{a} di Padova, vicolo Osservatorio, 3, 35122 Padova, Italy\\
$^{16}$Institute of Astronomy, University of Cambridge, Madingley Road, Cambridge CB3 0HA, UK\\
$^{17}$Department of Physics \& Astronomy, University of British Columbia, 6224 Agricultural Road, Vancouver, BC V6T~1Z1, Canada\\
$^{18}$ESO, Karl-Schwarzschild-Str. 2, 85748 Garching bei M\"unchen, Germany\\
$^{19}$UK Astronomy Technology Centre, Royal Observatory, Blackford Hill, Edinburgh EH9 3HJ, UK\\
$^{20}$Institut d'Astrophysique Spatiale (IAS), b\^atiment 121, Universit\'e Paris-Sud 11 and CNRS (UMR 8617), 91405 Orsay, France\\
$^{21}$Infrared Processing and Analysis Center, MS 100-22, California Institute of Technology, JPL, Pasadena, CA 91125, USA\\
$^{22}$School of Physics and Astronomy, The University of Manchester, Alan Turing Building, Oxford Road, Manchester M13 9PL, UK\\
$^{23}$Astronomy Centre, Dept. of Physics \& Astronomy, University of Sussex, Brighton BN1 9QH, UK\\
$^{24}$Institut d'Astrophysique de Paris, UMR 7095, CNRS, UPMC Univ. Paris 06, 98bis boulevard Arago, F-75014 Paris, France\\
$^{25}$Space Science \& Technology Department, Rutherford Appleton Laboratory, Chilton, Didcot, Oxfordshire OX11 0QX, UK\\
$^{26}$Institute for Space Imaging Science, University of Lethbridge, Lethbridge, Alberta, T1K 3M4, Canada\\
$^{27}$Centre for Astrophysics Research, University of Hertfordshire, College Lane, Hatfield, Hertfordshire AL10 9AB, UK\\
$^{28}$Harvard-Smithsonian Center for Astrophysics, 60 Garden Street, Cambridge, MA 02138, USA}}

\begin{document}

\date{accepted version $22^{\rm nd}$ December 2010}

\pagerange{\pageref{firstpage}--\pageref{lastpage}} \pubyear{2002}

\maketitle

\label{firstpage}

\clearpage
\begin{abstract}
We examine the rest-frame far-infrared emission from powerful radio sources 
with 1.4\,GHz luminosity densities of $25$$\le$$\log(L_{1.4}$/WHz$^{-1})$$\le$$26.5$ 
in the 
extragalactic {\it Spitzer} First Look Survey field. We combine 
{\it Herschel}/SPIRE flux densities with {\it Spitzer}/IRAC and MIPS 
infrared data to obtain total ($8-1000\,\mu$m) infrared luminosities for 
these radio sources. We separate our sources into a moderate, $0.4$$<$$z$$<$$0.9$,  
and a high, $1.2$$<$$z$$<$$3.0$, redshift sub-sample and we use {\it Spitzer} 
observations of a $z$$<$$0.1$ 3CRR sample as a local 
comparison. By comparison to numbers from the SKA Simulated Skies we find that our
moderate redshift sample is complete and our high redshift sample is 14\,per cent 
complete. We constrain the ranges of mean star formation rates (SFRs) to be 
$3.4$$-$$4.2$, $18$$-$$41$ and $80$$-$$581\,{\rm M}_\odot$yr$^{-1}$ for the 
local, moderate 
and high redshift samples respectively. Hence, we observe an increase in the 
mean SFR with increasing redshift which we can parameterise as $\sim(1+z)^Q$, 
where $Q=4.2\pm0.8$.
However we observe no trends of mean SFR with radio luminosity 
within the moderate or high redshift bins. 
We estimate that radio-loud AGN in the high redshift sample contribute
$0.1$$-$$0.5\,$per cent to the total SFR density at that epoch. 
Hence, if all luminous starbursts host radio-loud AGN we infer a radio-loud 
phase duty cycle of $0.001$$-$$0.005$.
\end{abstract}


\section{Introduction}

There is now strong evidence that powerful active galactic nuclei (AGN) 
played a key role in the evolution of galaxies. The correlation of central black 
hole and stellar bulge mass \citep{Magorrian:98}, and the increased prevalence
of star formation \citep{Hopkins:06, Giavalisco:04b} and AGN activity 
\citep{Wall:05,Aird:10} at earlier epochs
suggest that the growth of the black hole is somehow 
related to the growth of the host galaxy. In the local 
Universe we see little evidence of  high star formation rates (SFRs) in 
galaxies with powerful radio-loud AGN activity \citep[e.g][]{Condon:98b,Mauch:07}.
In the distant Universe, $z>1$, luminous
radio galaxies \citep{Seymour:07} and powerful 
starbursts \citep{Borys:05,Casey:09} are both hosted by massive galaxies, 
suggesting a common parent population.
The idea that these processes are likely to be connected at the epoch when black 
holes and galaxies went through their most rapid phases of growth
has been invoked
within various semi-analytical models in order to reconcile those models with 
observations \citep[e.g.][]{Springel:05}.

This connection between central black hole growth and star formation rate is 
often considered in the context of  `feedback' process(es), as the former is 
postulated to regulate the latter.
In particular there is observational evidence, as well as theoretical models, 
in which the jet from an AGN 
can produce either positive or negative feedback, where the jet, traced
by its radio emission, stimulates or quenches star formation respectively.
There is some observational evidence of {\em positive} feedback, whereby star 
formation is triggered by an AGN jet, e.g. in Minkowski's Object by a jet 
from NGC 541 \citep{vanBreugel:85,Croft:06}, as well as theoretical models 
which suggest that the shocks generated by jet propagation can trigger collapse 
of over-dense clouds and lead to star formation \citep{Fragile:04, Saxton:05}. 
{\em Negative} feedback by AGN jets would likely require the removal of fuel 
for star formation, evidence for which are the  powerful AGN-induced 
outflows which have been seen in high redshift radio galaxies
\citep{Nesvadba:06,Nesvadba:08}. Such a scenario has also been proposed to 
regulate the growth of massive galaxies in semi-empirical models 
\citep{Croton:06, Bower:06}, but this process is only important globally
at late times, $z<1$. At earlier times it would be most important
in halting the growth of the most massive galaxies.

Star formation in powerful AGN has been difficult to trace so far.
This difficulty is due to heavy contamination in traditional diagnostics by 
emission from the AGN (e.g. UV luminosity or optical emission line strengths) 
as well as obscuration by gas and 
dust. However the far-infrared (far-IR) presents a window 
in the electromagnetic spectrum where AGN emission is weak and star formation, 
if present, can dominate. AGN dust emission tends to peak in the near/mid-IR so 
far-IR emission should be a cleaner measure of SFR than other 
traditional methods. It is also possible to use the near/mid-IR to model and subtract
any potential AGN contribution to the far-IR \citep[e.g.][]{Hatziminaoglou:08}.

There is evidence for extreme SFRs in many powerful high redshift radio 
galaxies \citep[$z>2$, 1.4\,GHz luminosity densities, 
$L_{1.4}\ge10^{27}\,$WHz$^{-1}$,][]{Miley:08}
from their strong sub-mm emission 
\citep{Archibald:01, Reuland:04, Greve:06}, their mid-IR spectra 
\citep[][J. Rawlings, 2011, in prep.]{Seymour:08b} and the spectacular ($>$$100\,$kpc) 
Ly$\alpha$ haloes sometimes observed \citep{Reuland:03, VillarMartin:03} 
showing the extended gas that can provide the fuel for star formation. 
To compliment future targetted {\it Herschel\,} studies of the rare, 
very powerful radio-loud AGN, we examine in this work less luminous radio-loud 
AGN, $26.5>\log(L_{1.4}$/WHz$^{-1})\ge 25$, which can be found in reasonable 
abundance over areas of a few square degrees. We use this definition of
`radio-loud' AGN, based on radio luminosity density \citep[\eg][]{Miller:90},  
in order to avoid making any distinction between type 1 and type 2 AGN, 
i.e. AGN classification based upon optical spectroscopy, where different
amounts of AGN obscuration may affect the relative amount of optical emission.
Although, as we shall show, most of these sources are also `radio-loud' 
when using the definition of 
\citet[][5\,GHz over $B-$band luminosity $>10$]{Kellerman:89}.
Star formation in these less luminous radio-loud AGN remains poorly studied, 
as there has been no systematic follow-up of such sources above $z>0.1$.
Recently, the importance of radio-loud AGN in this luminosity range was 
demonstrated by \citet{Sajina:07} who found that $40\,$per cent of $z\sim2$
ULIRGs with deep silicate absorption features were radio-loud and  those authors
postulated that such sources are transition `feedback' objects after the 
radio jet has turned on, but before feedback has halted black hole accretion 
and star formation.

The SPIRE instrument \citep{Griffin:10} on board the {\it Herschel Space 
Observatory} \citep{Pilbratt:10} gives us a clear view of the 
far-IR/sub-millimeter Universe at wavelengths where many galaxies emit most 
of their luminosity. The Herschel Multi-tiered Extragalactic Survey 
(HerMES\footnote{http://hermes.sussex.ac.uk}, Oliver et al. 2011, in prep)
provides deep infrared SPIRE data over many of the best 
studied extra-galactic survey fields. Recent results from {\it Herschel} 
show that SPIRE detected AGN in deep HerMES fields have far-IR colours 
similar to the bulk of the SPIRE population which are believed to be star 
formation dominated
\citep{Elbaz:10, Hatziminaoglou:10} and  modeling of their  
spectral energy distributions (SEDs) suggests the SPIRE emission in AGN
is dominated by a star forming component \citep{Hatziminaoglou:10}.

The work presented here uses {\it Herschel}/SPIRE observations of the 
{\it Spitzer} Extragalactic First Look Survey (FLS) field taken as part of 
the {\it Herschel} 
Science Demonstration Phase (SDP) in October to November 2009. Of the fields 
observed in SDP this field had the best combination of wide area, uniform 
radio coverage and good multi-wavelength follow-up. We present our sample of 
moderate and high redshift radio-loud AGN in \S2 and, we derive IR
luminosities and star formation rates in \S3. We present our results in 
\S4 and discuss them in \S5. We conclude this paper in \S6.
Throughout we use a `concordance' cosmology of
$\Omega_{\rm M} = 1 - \Omega_{\Lambda} = 0.3$, $\Omega_0 = 1$, and 
$H_0 = 70\, \kmpspMpc$.

\section{Sample}

\subsection{Radio Sample and Cross Identification}
\label{sec:radio}

Our radio data come from the $1.4\,$GHz Very Large Array catalogue of 
\citet{Condon:03} which is complete down to $0.115\,$mJy ($5\,\sigma$).
We restrict our analysis to a region of the FLS with complete 
optical and near/mid-IR coverage, defined by $257.8^\circ<$~RA~$<261^\circ$ 
and $58.6^\circ<$~dec.~$<60.4^\circ$. These optical to mid-IR ancillary 
data were taken from the IRAC-selected, multi-wavelength data fusion 
catalogue in the FLS (hereafter the FLS `Data Fusion Catalogue') 
presented by Vaccari et al. (2011 in prep.).
The Data Fusion Catalogue is a {\it Spitzer}/IRAC-selected wide-area
multi-wavelength catalog covering the $\sim60\,\deg^2$ extragalactic fields 
covered by {\it Spitzer}/IRAC and MIPS $7-$band imaging. The main selection 
of the catalog requires an IRAC $3.6$ or $4.5\,\um$ detection, since the two 
{\it Spitzer} channels reach about the same depth. 
MIPS $24\,\um$ detections are associated with IRAC sources to improve
their positional accuracy, and the MIPS 70 and $160\,\um$ detections are 
confirmed by a MIPS $24\,\um$ detection to increase their reliability.

In this paper we use the version of the Data Fusion employed in HerMES 
SDP work. For the FLS field, we thus use the IRAC catalog from \citet{Lacy:05}, 
the MIPS $24\,\um$ catalogue from \citet{Fadda:06}, and MIPS 70 and $160\,\um$ 
catalogues produced by the HerMES team using the SSC provided software 
\citep[e.g.][]{Frayer:09}.
We combine the mid and far-infrared data from 
{\it Spitzer} with optical data ($ugriz$) from the Isaac Newton Telescope
(Solares et al., 2011, in prep.) as well as redshift information from the 
literature.

\begin{table}
 \centering
  \caption{Composition of FLS master radio catalogue. We indicate the total 
number of sources in the master catalogue, 
the number with cross-identifications in the FLS Data Fusion Catalogue and redshifts,
and the number of sources with redshifts and SPIRE/$250\,\um$ detections }
  \begin{tabular}{@{}lr@{}}
  \hline
  \hline
   Total number of radio sources & 1907 \\
  \hline
   with FLS Data Fusion XIDs and known redshifts & 885\\
   with SPIRE/$250\,\mu$m and known redshifts & 436\\
\hline
  \hline
\end{tabular}
\label{master}
\end{table}

The redshifts come from the Sloan Digital Sky Survey (SDSS) spectroscopy and 
photometry as well as dedicated follow-up of many radio and mid/far-infrared 
selected targets by several groups
\citep[\eg][]{MartinezSansigre:05,Papovich:06a,Weedman:06,Lacy:07,Marleau:07,Yan:07,Sajina:08,Dasyra:09}.
As the photometric redshifts from the SDSS do not extend accurately above $z=1$,
higher redshift sources will be dominated by the selection criteria 
of these different groups.
We can compare the optical magnitudes and mid-IR flux densities of the sources
with and without known redshifts. We find that around 100 radio sources 
with known redshifts are not detected in 
the $z-$band, but are detected at $24\,\mu$m at brighter flux densities than 
most sources without redshift information. Hence, as faint $z-$band sources 
typically lie at higher redshifts, then this observation is consistent 
with the specific targetting of bright $24\,\um$ sources for spectroscopic 
follow-up at high redshift.
We discuss how we deal with this selection in \S4.

   \begin{figure}
   \centerline{
   \psfig{file=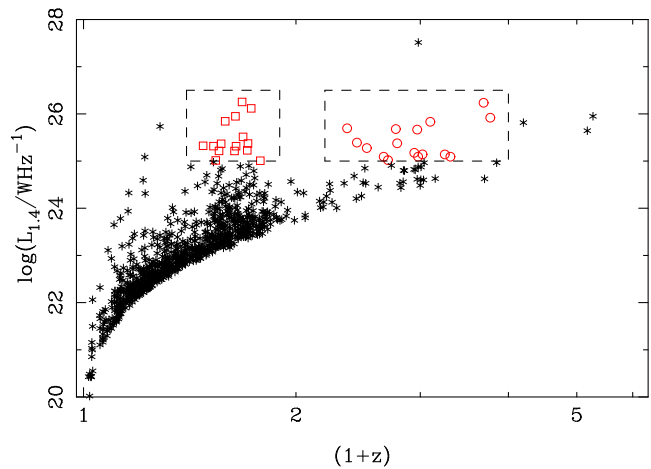,angle=-90,width=8.5cm}}	
      \caption{Redshift/radio luminosity distribution of 885/1907 radio sources 
in our master catalogue with known redshifts. The red symbols within the dashed 
rectangles indicate our moderate and high redshift sub-samples represented by 
squares and circles respectively. Note the sub-samples are chosen
in redshift ranges where they are likely to be most complete (see Fig.~\ref{fig2}).}
         \label{fig1}
   \end{figure}

We cross-correlated the radio catalogue
with the FLS Data Fusion Catalogue using a 2\,arcsec search 
radius between the radio and mid-IR ($3.6\,\um$) positions. 
Extended/multi-component sources from \citet{Condon:03} 
were inspected by eye and five were reclassified as being two or more 
separate sources due to the presence of more than one optical/near-IR 
counterpart to individual radio components.
We therefore obtained a master catalogue 
of 1907 radio sources 
of which 885 have spectroscopic or photometric redshifts from the Data
Fusion Catalogue (see Table~\ref{master}).
We illustrate in Fig.~\ref{fig1} the distribution in redshift/luminosity 
space of the sources from the master catalogue with known redshifts.
Our search radius and the
sky density of the FLS Data Fusion Catalogue imply that 12/1571 (i.e. $<1\,$per cent)
of our cross-identifications are by chance. 

While the redshift information for our sample is incomplete, it is only 
important for sources that potentially satisfy our radio luminosity selection 
criteria and are hence included in our radio-loud sample. 
However, in the subsequent sections we present the selection of our 
radio-loud AGN samples in two different redshift ranges, assess how 
complete these are by comparisons to models based on the known evolution of 
the high redshift radio-loud population, see \S2.3, and how this selection will 
effect our sample, see \S4.

\subsection{Radio-loud Selection \& Sub-samples}

To obtain accurate luminosities, radio spectral indices are required, so we 
cross-correlated the master catalogue with the $610\,$MHz catalogue of 
\citet{Garn:07} finding counterparts within $6\,$arcsec for $68\,$per cent of the
master sample. We use a $6\,$arcsec search radius to account for the positional
accuracy of the 610\,MHz data. For radio 
sources without $610\,$MHz counterparts we assumed a spectral index 
with a value of $\alpha=-0.75$ ($S_\nu\propto \nu^\alpha$) consistent with the
mean value found for faint radio sources in general
\citep[AGN and starbursts alike e.g.][]{Ibar:09}. We note that the sample here 
has a slightly
steeper mean radio spectral index ($\alpha=-0.82$), but the relative limits of 
the 1.4\,GHz and 610\,MHz survey result in bias against sources with a flat 
spectrum at low flux densities.
We select our radio-loud AGN sample with luminosity density cuts of 
$25\le\log(L_{1.4}/$WHz$^{-1})\le26.5$. The lower limit is chosen to ensure
our sources are genuinely radio-loud and to minimise the number of extreme 
star forming galaxies (SFG) selected. Indeed, this lower radio luminosity 
is equivalent to a total IR ($8-1000\,\mu$m) luminosity of 
$\sim3\times10^{13}\,L_\odot$ from the
correlation of far-IR and radio luminosities for star forming galaxies 
\citep{Yun:02} and therefore a SFR of $\sim6000\,{\rm M}_\odot$yr$^{-1}$ using the
relations of \citet{Kennicutt:98}. Hence, this luminosity would be extreme 
for a starburst galaxy. The upper limit is imposed as radio 
sources with luminosities greater than this cut are rare in the volume 
probed in this study.
We find one source with such a luminosity 
($L_{1.4}\sim10^{27.5}\,$WHz$^{-1}$ at $z\sim2$, see Fig.~\ref{fig1}) which is 
identified as a SDSS QSO. We consider it no further in this study, but note
that this radio-loud QSO is not detected in our SPIRE 
observations. We also find that all our `radio-loud' AGN would
also be classified as radio-loud by the rest-frame $5\,$GHz to $B-$band flux
ratio according to the criteria of \citet{Kellerman:89} bar three sources
in the high redshift bin which have ratios just below the cut-off value of
ten.

We then separate the luminous radio sources into moderate 
($0.4< z<0.9$) and high ($1.2< z<3$) redshift samples with 15 and 16 
sources respectively (out of a total of 36 radio sources from the master
catalogue with $25<\log(L_{1.4}/{\rm WHz}^{-1})<26.5$). 
We chose these two redshift bins since the redshift distribution of the 
luminous radio sources peaks in these ranges 
(see Fig.~\ref{fig1}) and hence we should obtain the most complete sub-samples 
possible given the data available (see below for estimates of their completeness).  
We note that the general decrease 
in known redshifts at $z\sim1$ seen in Fig.~\ref{fig1} is due to the 
ineffectiveness of SDSS photometric and spectroscopic redshift estimation above 
this redshift. Hence, all the sources in the moderate redshift sample have redshifts 
from SDSS (4/15 are spectroscopic with ther remainder being photometric). 
Sources with higher known redshifts are generally from targetted 
follow-up of various classes of object as well as the occasional SDSS QSO.
All the redshifts in the high redshift bin are spectroscopic and come from 
these various follow-up projects. Interestingly, these two 
redshift ranges also cover similar length cosmic epochs of about $3\,$Gyr each.
The median radio 
luminosities of both sub-samples are very similar: 
$\log(L_{1.4}/{\rm WHz}^{-1})=24.9$ and $25.0$ for the moderate and high redshift 
samples respectively.

   \begin{figure}
   \centerline{
   \psfig{file=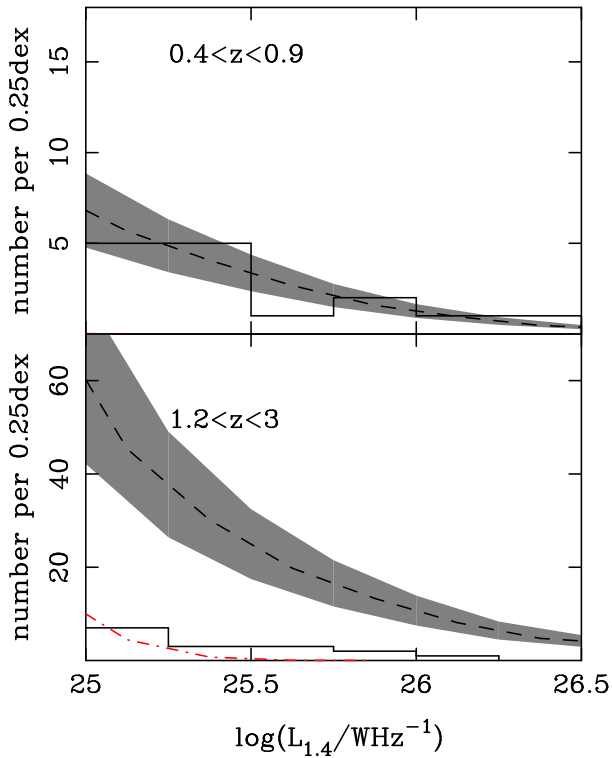,angle=-90,width=8cm}
}	
      \caption{Observed number distribution versus radio luminosity density 
	of sources 
	in our moderate and high redshift samples (black solid histograms).
	The dashed line represents the distribution of 
	the total number of radio sources
	expected in this volume from the 
	SKA Simulated Skies (S-cubed, Wilman et al. 2008), where the number 
	expected to be SFGs is indicated by the dot-dashed line (none are 
	predicted in the moderate redshift sample). 
	The shaded region represents a $30\,$per cent uncertainty in S-cubed.
	In comparison to S-cubed, our moderate redshift sample is $100\,$per cent
	complete and our high redshift sample is $14\,$per cent complete.}
         \label{fig2}
   \end{figure}

\subsection{Completeness}
\label{comp}

In Fig.~\ref{fig2} we show the {\em observed} distribution of  radio 
luminosities in each redshift sample, and compare them to the {\em modeled} 
luminosity distributions over the same volume derived from the SKA 
Simulated Skies \citep[S-cubed,][]{Wilman:08} at the radio flux density limit
of the FLS ($0.115\,$mJy). As well as the total number
of sources predicted in these luminosity redshift bins, we also indicate the 
number of extreme SFGs (SFR$\,>6000\,{\rm M}_\odot$yr$^{-1}$) predicted.
The class of AGN from S-cubed which dominate this distribution are the 
low-luminosity radio-loud AGN \citep{Wilman:08}. The evolution of this
population is taken from `model C' of \citet{Willott:01} and is 
reasonably well constrained up to $z=2$. We then apply a high redshift
decline in space density represented by $(1+z)^{-2.5}$ above $z=2.5$
as recommended in \citet{Wilman:08}.
There is also a small, $\sim6\,$per cent, contribution to the number of
sources predicted by S-cubed of `radio-quiet' AGN whose evolution is 
less well constrained by observation.
We have included a $30\,$per cent uncertainty in the predicted number of radio 
sources from S-cubed to represent the uncertainty in the evolution of the 
luminosity function for the low luminosity radio-loud AGN population, in 
particular the high redshift cutoff and the less
well constrained `radio-quiet' population, as well as 
sample variance for a survey field covering only a few square degrees.

We find that our moderate redshift sample is complete given the uncertainties
we ascribed to S-cubed. However, we
find that the number of sources predicted by S-cubed exceeds the 
number we observe in the high redshift bin implying a 14\, per cent completeness. 
The numbers of sources we find in each sub-sample compared to the number
predicted from S-cubed is given in Table~\ref{subs}.
The number of sources deficient in our high redshift sample (and at other 
redshift ranges) can be accounted for by the lack of redshift information in 
the master sample (Table~\ref{master}). 
We account for any bias in our samples, e.g. mid-IR selection of known high 
redshift radio-loud AGN,  in \S4 by considering the 
full limits of the completeness and we demonstrate that we 
can still constrain the range of mean SFRs for these 
samples by making two extreme assumptions about the sources 
missing from our sample.

\section{Analysis}

\subsection{IR Luminosities}

\begin{figure}
\centering
\psfig{file=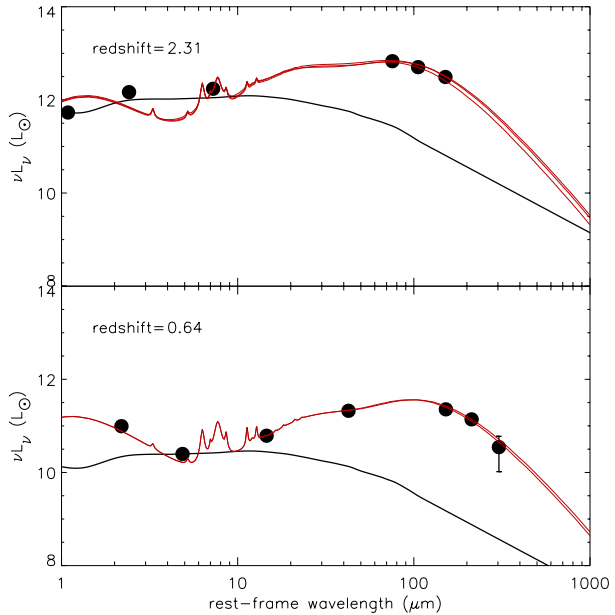,angle=0,width=8.0cm}
   \caption{Example SED fits to the available IR photometry from $3.6$ to 
    $500\,\um$ where we show rest-frame luminosity plotted against rest-frame 
    wavelength. The red lines indicate the best fit starburst template and the 
    range of templates within $\Delta\chi_{i}^2<1$ and the black lines indicate
    the maximum normalisation of the AGN template to the lowest mid-IR photometry.
    The filled circles indicate the {\it Spitzer} and {\it Herschel} photometry 
    used in the fitting. Note 
    that in most cases the uncertainties are smaller than the symbols. 
    We present an object from both the low redshift sample (lower panel, $z=0.645$, 
    $L_{IR}=5.77\pm0.58\times10^{11}\,L_\odot$) and the high redshift 
    sample (upper panel, $z=2.31$, $L_{IR}=1.84\pm0.17\times10^{13}\,L_\odot$). }
      \label{fig3}
\end{figure}

We extracted SPIRE flux densities at the positions of all radio and 
$24\,\um$ sources 
using the HerMES XID method \citep{Roseboom:10}. This approach minimizes the 
effect of source blending, as the SPIRE flux densities are estimated via 
linear inversion methods using the positions of known $24\,\um$ sources, or 
radio position if there is no $24\,\um$ counterpart, as a prior. 
In Roseboom et al. 2010 the $250\,\um$ flux density 
uncertainty is estimated to
be $7.45\,$mJy from injection and recovery of mock sources into the observed 
maps. Flux density uncertainties are obtained from the RMS of input-output
flux densities and consequently include contributions from both instrumental 
and confusion noise.
We find 436 sources having $250\,\mu$m counterparts with $>3\,\sigma$ 
detections and known redshifts. 
For the radio-loud sources we find 4/15 and 9/16 with significant
$250\,\um$ detections in the moderate and high redshift bins 
respectively\footnote{For reference, $24\,$per cent of the radio sources with 
unknown redshifts have significant detections in the SPIRE wavebands}.

The relative depths of the $24\,\um$ and $250\,\um$ data available for this
field have some bearing on the how the $250\,\um$ sources are found by the 
XID method.
The $24\,\mu$m imaging data of this field is relatively shallow 
with respect to the SPIRE data (compared to other fields to 
which this method has been applied).
Hence, there  may be non-negligible 
$250\,\mu$m flux remaining in the field which has not been extracted due to the 
lack of a $24\,\mu$m (or radio) counterpart.
We visually inspected all $250\,\um$ detections of the radio-loud sources 
in the SPIRE image and they all appear isolated with no sources close enough to 
them which could significantly effect the measurement of their SPIRE flux density.

We derive total ($8-1000\,\um$) IR luminosities by fitting all the data 
available for the 436 radio sources
across the {\it Spitzer}/IRAC$+$MIPS and 
{\it Herschel}/SPIRE bands following the method outlined in 
\citet[][see Fig.~\ref{fig3}]{Symeonidis:09}. 
In all cases we use the {\it Spitzer}/$24\,\um$
and {\it Herschel}/$250$/$350$/$500\,\um$ photometry although in some
cases the $350$ and $500\,\um$ photometry have extremely large uncertainties
due to  their low SNR, $<3$ and do not significantly affect the values of $\chi^2$
dervived. 
This fitting method uses all the models from 
\citet[][]{Siebenmorgen:07}, which cover a wide range of SED types, 
and finds the best fit using standard $\chi^2$
minimisation from which a total IR luminosity is calculated. Uncertainties 
in the IR luminosity are derived from the range of values obtained from SED 
fits which differ from the best fit by 
$\Delta\chi_{i}^2=(\chi_i^2 - \chi_{\rm min}^2)<1$.

\subsection{AGN contribution to the far-IR luminosity}
\label{agnc}

A further issue to consider, if we are to use the total IR luminosities 
as indicators of SFR, is the AGN contribution to this luminosity which could 
lead to an over-prediction of the star formation rates. This issue is 
especially important because our sources are selected to be AGN.
In a similar fashion to \citet{Symeonidis:10}, we address this issue 
by normalising a QSO template from \citet{Elvis:94} to the data point 
with the lowest luminosity from our photometric dataset of $3.6-24\,\mu$m, 
as the AGN emission must be constrained by our photometry. If we use other 
AGN SED models \citep[e.g. type 1 and type 2 AGN from][]{Polletta:07}
we find that our estimates of the upper limits to the AGN luminosities and 
ratios of AGN to total IR luminosities change little, $<10\,$per cent (and 
therefore even less for the final SFR). Such model SEDs are broadly similar 
to the \citet{Elvis:94} templates which in the IR are generally flat (in 
$\nu L_\nu$) out to the far-IR where they then drop sharply. 
We note that only four of our high redshift sources
have mid-IR spectrsocopy from \citet{Yan:07}, hence we do not use 
these data to constrain SED fits.

\begin{figure}
\centering
\psfig{file=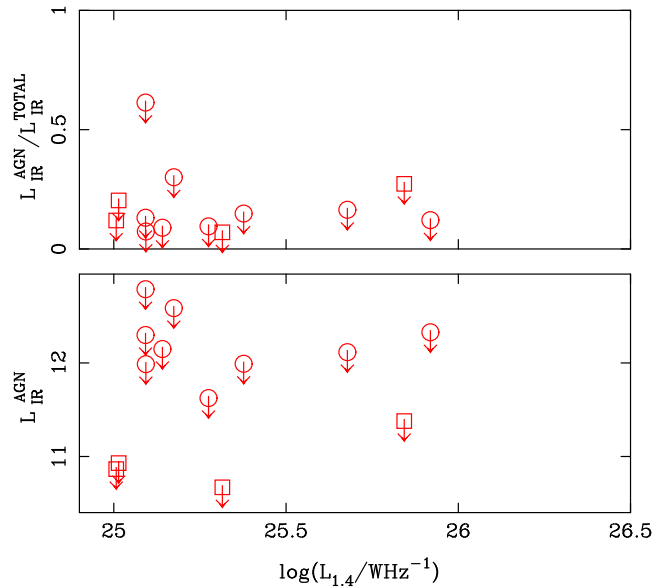,angle=-90,width=8.5cm}
   \caption{Upper limits to the AGN IR luminosity estimated from the 
     normalisation of an AGN SED to the lowest mid-IR luminostiy, and the 
     resulting upper limit to the ratio of AGN to total IR luminosity, both 
     plotted as a function of radio luminosity density. These sources are from 
     both our moderate and high redshift samples (open squares and circles 
     respectively).}
      \label{fig5}
\end{figure}

We then estimate the AGN contribution to the total infrared luminosity by 
integrating the QSO template in 
the $8-1000\,\um$ region and subtract this from our total infrared luminosity
to obtain a star foming IR luminosity for each object. 
We can then convert this star forming IR luminosity to a
SFR using the \citet{Kennicutt:98} relation. 
In Fig.~\ref{fig5} we show the AGN IR luminosity and ratio of 
AGN to total IR luminosity as a function of radio luminosity density 
for the radio sources detected by SPIRE in our two redshift samples. 
The IR AGN luminosity has a large scatter which is largely due to the moderate
redshift sub-sample having lower IR AGN luminosities ($\sim10^{11}\,L_\odot$)
than the high redshift sub-sample ($\sim10^{12}\,L_\odot$), although we 
observe no trend with radio luminosity within a sub-sample. 

We suspect that the greater AGN IR luminosities of the sources in the 
high redshift sample is most likely due to a bias in the redshift 
identification toward sources with bright $24\,\um$ flux densities 
($\ge 1\,$mJy) as discussed in \S2.1. Additionally, the flux 
limited nature of the {\it Spitzer} and {\it Herschel} data mean that
the SPIRE observations are 
more sensitive to lower IR luminosities at lower redshifts. 
As these indentified sources comprise just $14\,$per cent of the high 
redshift sample they are not likely to be representative in terms of 
their AGN fraction.

The ratio of AGN to total IR luminosity tends to be low, under 0.3 bar one 
source, consistent with the results seen in \citet{Hatziminaoglou:10}, and 
averages around $0.15$. As a check we apply the simultaneous AGN/starburst 
template 
fitting routine used by \citet{Hatziminaoglou:10} to the radio-loud AGN 
studied here and we find similar total IR luminosities and AGN fractions.
Therefore, the final SFRs we derive are not very sensitive to our choice of 
model starburst and AGN SEDs. Our assumption 
that the mid-IR is completely dominated by the AGN, while conservative, also 
does not have a strong effect on the final SFR due to the low AGN fraction.

\begin{figure}
\centering
\psfig{file=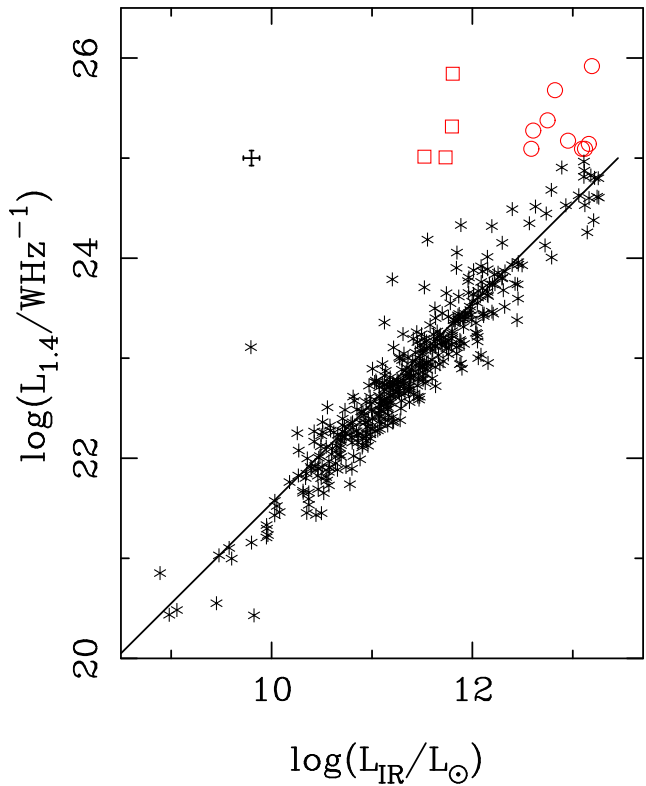,angle=-90,width=8.5cm}
   \caption{Total infrared luminosity plotted against radio luminosity for 
	all the FLS radio sources with redshifts and a $>3\,\sigma$ detection 
	at $250\,\um$ (including confusion noise). The sample 
	of luminous radio sources, $L_{1.4}>10^{25}\,$WHz$^{-1}$, used in this 
        work are indicated by open symbols and the less luminous radio sources 
	by asterisks. Note the IR luminosities of the luminous radio sources
	do not include the AGN contribution (see \S3.3). 
	We fit the observed correlation in luminosities for 
	$L_{1.4GHz}<10^{24}\,$WHz$^{-1}$ and $L_{IR}<10^{12.5}\,L_\odot$ and 
	derive $q_{IR}= 2.40 \pm 0.19$. Two radio-loud AGN lie above, but within 
	$2\,\sigma$ of  this correlation, however they have radio
	spectral indices inconsitent with star formation. Hence we conclude
	that their radio emission is dominated by AGN processes.}
      \label{fig4}
\end{figure}

\subsection{Comparison between radio and IR luminosities}
\label{firrc}

We calculate the total IR luminosities of all SPIRE detected sources in order to 
confirm our method of measuring these luminosities by comparison with the
radio/far-IR correlation seen in local star forming galaxies and now confirmed 
at higher redshifts \citep[][]{Seymour:09,Ivison:10}.
Additionally, by extrapolating this empirical correlation to higher 
luminosities we can assess the contribution of star formation to brighter 
radio sources. Our radio luminosity selection would be equivalent to a 
SFR of $\sim6000\,{\rm M}_\odot$yr$^{-1}$ for a pure SFG, but potentially 
there could be a few sources with higher SFRs within the volume probed here 
(see S-cubed predictions in Fig.~\ref{fig2}).

We find a strong correlation between the radio and IR luminosities, 
particularly below $L_{1.4}=10^{24}\,$WHz$^{-1}$ and 
$L_{\rm IR}=10^{12.5}\,L_\odot$ (see Fig.~\ref{fig4}) which we use to verify our 
IR luminosities. Note the IR luminosities of the radio sources in the 
moderate and high redshift samples have had the AGN contribution removed.
We define the ratio of radio to IR 
luminosity as $q_{\rm IR}=\log(L_{\rm IR}/L_{1.4})+14.03$ 
\citep[as used in][this definition is an equivalent, but more 
convenient form than the classical one of Helou et al., 1985]{Sajina:08}.
\nocite{Helou:85} By fitting the
correlation over these luminosity ranges we get a value of 
$q_{\rm IR}=2.40\pm 0.19$ using a biweight estimator \citep{Beers:90},
in good agreement with the value found locally \citep{Yun:02} and at higher 
redshift \citep{Ivison:10}.

We then assume that this relation holds to higher luminosities (i.e. to 
SFRs $>6000\,{\rm M}_\odot$yr$^{-1}$) and note that 2 radio sources with 
$\log(L_{1.4}/$WHz$^{-1})>25$ lie just within $2\,\sigma$ of $q_{\rm IR}$
The proximity of these 2 sources to the correlation 
may mean that these sources have a non-negligible contribution of 
star formation to their radio luminosity. The AGN fraction of the total
IR luminosities for these sources is low, $\le 10\,$per cent.
However, upon closer inspection 
these two sources have radio spectra which are either too steep, 
$\alpha_{1.4}^{610}=-1.78$, or too flat, $\alpha_{1.4}^{610}=-0.22$,
compared to the canonical value for star forming galaxies \citep{Condon:92}. 
The fraction of the radio luminosity due to star formation, assuming the
star forming component lies precisely on the correlation, is $25-30\,$per cent.
Hence, we conclude that their radio emission is dominated by AGN processes 
and retain them within our high redshift sample.

\subsection{Stacking the non-detections at $250\,\um$ }

We can obtain an approximate constraint on the far-IR luminosity of the radio-loud
AGN not detected at $250\,\um$ in each sample by employing stacking techniques to 
obtain mean $250\,\um$ flux densities for those sources. By assuming the same 
distribution of redshifts, IR SED types and ratios of AGN to total IR luminosity, 
we can argue that the mean SFRs of the undetected and detected samples scale 
directly with the $250\,\um$ flux densities within both redshift ranges.
Therefore we stacked the 11 and 7 sources not detected at $250\,\um$ in each 
sub-sample and find the mean flux densities reported in Table~\ref{subs}. The 
uncertainties in flux densities of the stacked sources are simply those of the mean.

\begin{table}
 \centering
  \caption{Composition of radio-loud AGN sub-samples. For both redshift sub-samples 
we present the number of sources predicted from S-cubed, the total number found, 
the number with SPIRE/$250\,\um$ detections, the mean $250\,\um$ flux densities 
of the detected and undetected (via stacking techniques) sources, the mean SFR of 
the detected sources, the inferred mean SFR of the undetected sources (assuming 
it scales directly with the 
mean $250\,\um$ flux density), the total mean SFR of all observed sources and the 
range of mean SFRs given the number of sources predicted by S-cubed.}
  \begin{tabular}{@{}lccc@{}}
  \hline
  \hline
   description  &  & ~~~~~~~sub-sample$\!\!\!\!\!\!\!\!\!\!\!\!\!\!\!\!\!\!\!\!\!$ & \\
                &  & moderate   & high \\
  \hline
\hline
   S-cubed number predicted  & & 16 & 116 \\
   total number found & & 15 & 16 \\
   with SPIRE/$250\,\mu$m $\sigma\ge3$ & & 4 & 9 \\
  \hline
   $\left\langle S^{\rm detected}_{250}\right\rangle$ & $\!\!\!\!\!\!\!\!\!\!$(mJy) & $27.2\pm2.5$ & $39.2\pm2.5$\\
   $\left\langle S^{\rm undetected}_{250}\right\rangle$ & $\!\!\!\!\!\!\!\!\!\!$(mJy) & $2.0\pm0.8$ & $6.5\pm1.0$ \\
  \hline
   $\left\langle {\rm SFR}^{\rm detected}_{250}\right\rangle$ & $\!\!\!\!\!\!\!\!\!\!$(${\rm M}_\odot$yr$^{-1}$) & $92\pm28$ & $914\pm274$ \\
   $\left\langle {\rm SFR}^{\rm undetected}_{250}\right\rangle$ & $\!\!\!\!\!\!\!\!\!\!$(${\rm M}_\odot$yr$^{-1}$) & 6.7$\pm2.8$ & 153$\pm23$\\
   $\left\langle {\rm SFR}^{\rm total}_{250}\right\rangle$ & $\!\!\!\!\!\!\!\!\!\!$(${\rm M}_\odot$yr$^{-1}$) & $29.5\pm11.6$ & $581\pm143$ \\
  \hline
   range of $\left\langle {\rm SFR}\right\rangle$ &  $\!\!\!\!\!\!\!\!\!\!$(${\rm M}_\odot$yr$^{-1}$) & $18-41$ & $80-581$\\
\hline
\hline
\end{tabular}
\label{subs}
\end{table}

\subsection{SFRs in local ($z<0.1$) radio-loud AGN}

   \begin{figure*}
   \centering{
   \hskip 0.01cm
   \psfig{file=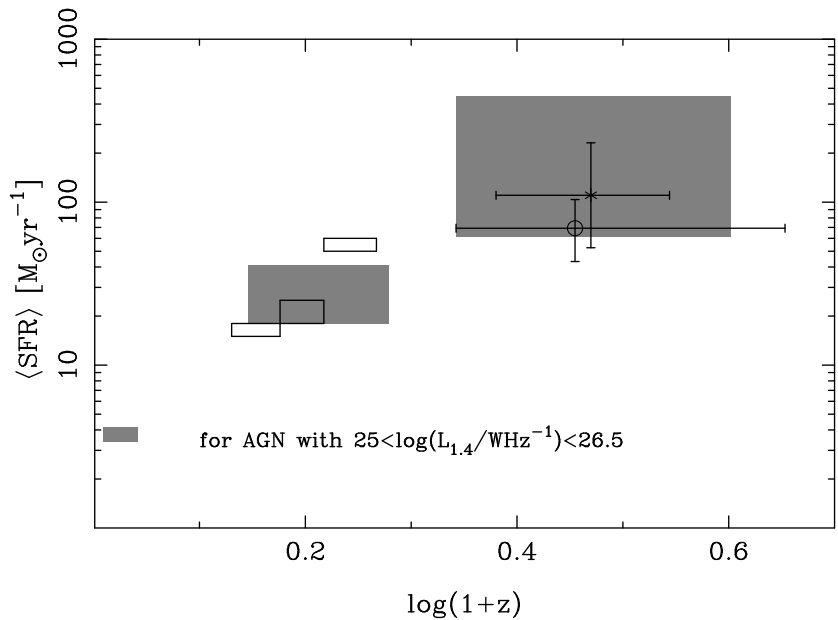,angle=-90,width=11cm}
}	
      \caption{Range of mean SFRs plotted as a function of redshift for radio-loud 
        AGN with $25\le \log(L_{1.4}$/WHz$^{-1})\le 26.5$ (shaded regions).
	At $0$$<$$z$$<$$0.1$ the data is from our 3CRR local reference sample and 
	at $0.4$$<$$z$$<$$0.9$, and  $1.2$$<$$z$$<$$3.0$ from our moderate and 
	high redshift sub-samples respectively. The open rectangles indicate 
	the results from 
	Hardcastle et al., (2010) using {\it Herschel} observations of sources 
	with a similar range of radio luminosities. The points with error-bars
	present the approximate mean SFRs of X-ray selected AGN over the range of 
	redshifts indicated from Lutz et al. (2010, open circle) and Shao et al. 
	(2010, asterisk).}
         \label{fig6}
   \end{figure*}

In order to examine any evolution of the mean SFR of radio-loud AGN over cosmic time
we need a local baseline to compare with. Recently published  {\it Spitzer}/MIPS 
observations of the local ($z<0.1$) 3CRR sample \citep{Dicken:10} provide an 
excellent opportunity to assess star formation in the nearby radio-loud population. 
The $1.4\,$GHz luminosity densities of this sample, derived from the $5\,$GHz 
values in \citet{Dicken:10} assuming $\alpha=-0.75$ and $S_\nu\propto\nu^\alpha$, 
fall within the $25<\log(L_{1.4}/{\rm WHz}^{-1})<26.5$ range of our sub-sample 
selection. The 3CRR sources were selected to only include sources with 
Fanaroff-Riley class II morpholgies \citep[i.e. those with radio lobes which are 
brightest at their edges,][]{Fanaroff:74}. However the lower radio luminosity 
density limit used in our work very closely corresponds to the luminosity density, 
$\log(L_{1.4}/{\rm WHz}^{-1})=25.1$,  at which the radio-loud population switches 
from mostly containing class I sources to mostly containing class II sources.
Furthermore, this local sample is not sensitive to the low end of our radio 
luminosity density range at $z=1$, and therefore may not be $100\,$per cent 
complete. Dicken et al. (2010) derive rest-frame $70\,\um$ luminosities from 
their {\it Spitzer}/MIPS observations which they compare with the [OIII] emission 
line luminosities of the local 3CRR sample. They find a broad correlation implying 
that generally the $70\,\um$ luminosity
is due to the AGN. However some 3CRR sources, which show evidence of star formation 
from their optical spectra, generally lie above this correlation, i.e. they have 
an excess of $70\,\um$ luminosity compared the [OIII] emission. These authors 
postulate that this $70\,\um$ excess could be due to star formation.

Here, we estimate the range of mean SFR in this sample using two assumptions. To 
obtain an upper limit we assume that {\em all} of the $70\,\um$ luminosity is due 
to star formation. To obtain a lower limit we use the linear regression fit by 
\citet{Dicken:10} to the correlation of the OIII and $70\,\mu$m luminosities to 
estimate the AGN only $70\,\um$ luminosity.
We then subtract the AGN luminosity from the total $70\,\um$ luminosity for
all sources lying more than $0.3\,$dex above the correlation in order to obtain 
a starburst only $70\,\um$ luminosity. In both cases we convert
the $70\,\um$ luminosities to total IR luminosities using the relation of 
\citet{Symeonidis:08} and then to SFRs using the Kennicutt (1998) relation as 
before. Due to the size of the sample and the influence of one very luminous 
source we use the median inferred SFR and find that the range of typical SFRs 
for the local 3CRR sample is $3.4-4.2\,{\rm M}_\odot$yr$^{-1}$ from 
these two assumptions.

  \section{Results}
  \label{sec:res}

In Table~\ref{subs} we report the mean SFR, $\left\langle{\rm SFR}\right\rangle$, 
of the radio-loud AGN detected at $250\,\um$ in each sub-sample.  
The SFRs of individual sources are derived from the total
IR luminosities, minus the AGN contribution (see \S~\ref{agnc}), using the 
conversion factors of \citet{Kennicutt:98}. We find 
values of $92\pm28\,{\rm M}_\odot$yr$^{-1}$ and $914\pm274\,{\rm M}_\odot$yr$^{-1}$ 
in the moderate and high redshift bins respectively. 
For the sources undetected at $250\,\um$ we find stacked $250\,\um$ flux densities 
which are a factor eleven and seven lower than the mean flux densities of the 
detected sources (see Table~\ref{subs}) for the moderate and high redshift 
sub-samples respectively. It is unsurprising that undetected sources have a mean 
flux density lower than those detected, but the fact they are considerably lower 
(i.e. not just below our $3\,\sigma$ cut) suggests that these radio-loud AGN have 
a wide range of intrinsic SFRs. We report the SFRs of the undetected sources in 
Table~\ref{subs} obtained from the ratio of the mean $250\,\um$ flux densities 
of the detected and undetected sources and the 
measured SFR of the detected sources.
Then we estimate the total mean SFR in each subset by combining the mean SFR of
the detected and undetected sources weighted by the number in each group.
The estimated total mean 
SFRs for the total sample are therefore $29.5\pm11.6\,{\rm M}_\odot$yr$^{-1}$ and 
$581\pm143\,{\rm M}_\odot$yr$^{-1}$ for the moderate and high redshift bins 
respectively.

The moderate redshift sample is complete within the uncertainties of the S-cubed 
simulation (we find 15/16 predicted sources in this redshift/luminosity density 
paramter space). Hence, we can directly calculate the mean SFR of the low redshift 
sample by summing the observed SFRs and dividing by the number sources. The 
uncertainties are simply those of the measured SFRs, which directly come from the 
uncertainties  in the IR luminosities, combined with the $30\,$per cent uncertainty 
in the S-cubed model. As the latter are so much greater than the former our 
uncertainties are dominated by the conservative uncertainties we used in S-cubed. 
We find a mean SFR for this sub-sample of $29.5\pm11.6\,{\rm M}_\odot$yr$^{-1}$ 
which is equivalent to the range of values of $18-41\,{\rm M}_\odot$yr$^{-1}$.

As we saw from Fig.~\ref{fig2} the high redshift sub-sample is incomplete, 
although we can quantify the incompleteness from comparisons to the S-cubed 
simulation. The number of radio-loud AGN expected from S-cubed is given in 
Table~\ref{subs}. We cannot estimate the properties of sources not included in our 
high redsift sample due to lack of redshift information. However, we can estimate 
likely lower and upper limits on the mean SFR from two simple assumptions. Firstly, 
to estimate the lower limit we assume that all the sources missed have SFRs of 
zero and then scale the mean SFR by the incompleteness (i.e. the lower limit is 
$\frac{16}{116}\,\times$ the mean SFR for the observed fraction). While $24$\,per 
cent of the sources with unknown redshifts have $250\,\um$ detections we have no 
way of knowing how many of these fall into our high redshift sub-sample, hence 
this method of determining our lower limit is the most robust approach. Secondly, 
for the upper limit we assume that all sources not included have mean SFRs 
identical to the detected fraction, i.e. the upper limit is simply the measured 
mean SFR for the detected fraction. Hence, 
we calculate the range of mean SFRs for the high redshift sub-sample to be 
$80-581\,{\rm M}_\odot$yr$^{-1}$.

   \begin{figure}
   \hspace {0.01cm}
   \psfig{file=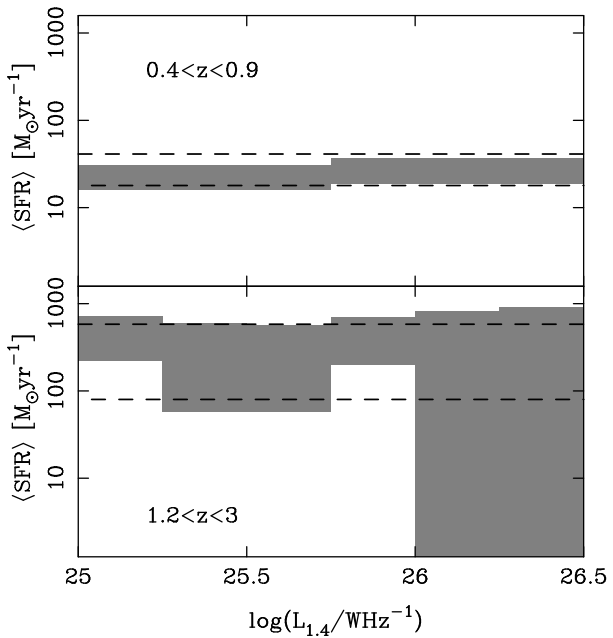,angle=-90,width=8cm}
      \caption{Mean star formation rate, $\left\langle{\rm SFR}\right\rangle$,
	of the host galaxies of radio-loud AGN plotted as a function 
	of radio luminosity density for each of the redshift sub-samples. 
	The shaded regions for each bin then represent the range 
	of range of $\left\langle{\rm SFR}\right\rangle$ assuming either 
	(i) all the sources missed have $\left\langle{\rm SFR}\right\rangle=0$
	(the lower limit) or (ii) all the sources missed have 
	$\left\langle{\rm SFR}\right\rangle$ equal to the sources found.
	The dashed lines represent the range of 
	$\left\langle{\rm SFR}\right\rangle$ 
	for the whole of each redshift sample as given in table~\ref{subs}.}
         \label{fig7}
   \end{figure}

We compare these constraints with those found for the local 3CRR sample and the recent
results of Hardcastle et al. (2010) in Fig.~\ref{fig6} who measure IR luminosities
from {\it Herschel}-ATLAS observations of radio sources occupying a
similar region of redshift/luminosity parameter space. We see an increase in 
the mean SFR of radio-loud AGN with cosmic look back time. In the local Universe
we found the mean SFR of $z<0.1$ radio-loud AGN to be 
$3.4-4.2\,{\rm M}_\odot$yr$^{-1}$, whereas at moderate redshifts, $0.4<z<0.9$, 
we constrain it to be $\sim5-10\,$times greater and in our high redshift 
sample we find it to be $\sim20-150\,$ times greater. While these ranges 
of mean SFRs are wide we observe a clear trend of increasing mean 
star formation rate with redshift in radio-loud AGN in the luminosity density 
range $25<\log(L_{1.4}/{\rm WHz}^{-1})<26.5$, a trend that is also seen over
a smaller redshift range in the results of \citet{Hardcastle:10}.

We can quantify this rate of increase by fitting a straight line 
through the shaded regions of Fig.~\ref{fig6} via linear regression. We then 
find that the mean SFR of radio-loud AGN in this luminosity range evolves 
as $(1+z)^Q$, where we measure the value of $Q=4.2\pm0.8$. This value for the 
evolution is strong and greater than that measured for the evolution of the 
star forming luminosity function \citep[which typically has values of $Q\sim3$ 
as traced by IR surveys][]{LeFloch:05,Huynh:07,Magnelli:09,Rodighiero:10}.
We can also compare our results with the mean SFRs of high redshift AGN 
selected at other wavelengths. The mean SFRs of X-ray selected AGN, 
$L_{2-10keV}>10^{43}\,$erg\,s$^{-1}$, have been 
studied recently by \citet{Shao:10} and \citet{Lutz:10} who find that such 
sources have mean SFRs within, but at the low-end of, the range 
of values found our high redshift bin. 
We illustrate those results in Fig.~\ref{fig6} using the same Kennicutt total IR 
luminosity to SFR conversion as before and converting the Shao et al. $60\,\um$
monochromatic luminosities using the formula presented in \citet{Symeonidis:08}.
Also, \citet{Hatziminaoglou:10} find a similar range of SFRs for a heterogenous 
sample of AGN above $z=1$, suggesting that this increase is common to
different types of AGN activity.

If we sum the observed star formation in each redshift bin we can calculate  
the comoving star formation rate density due to the host galaxies of the 
radio-loud AGN in each redshift sub-sample. We find values of 
$\sim2.5\times10^{-5}\,{\rm M}_\odot$yr$^{-1}$Mpc$^{-3}$ for the moderate 
redshift bin and  $1-5\times10^{-4}\,{\rm M}_\odot$yr$^{-1}$Mpc$^{-3}$
for the high redshift bin. For the local redshift bin, the star formation 
density due to the host galaxies of the radio-loud AGN 
is $\sim4\times10^{-8}\,{\rm M}_\odot$yr$^{-1}$Mpc$^{-3}$.
We can compare these SFR densities with the globally measured SFR history 
from a variety of different methods \citep[e.g.][]{Hopkins:06}. 
We observe that the relative contribution of the host galaxies of radio-loud 
AGN to the total comoving SFR density increases with redshift from $
\sim0.0004\,$per cent in the local sample to $\sim0.03\,$per cent and 
$\sim0.1-0.5\,$per cent for the moderate and high redshift samples respectively.

In Fig.~\ref{fig7} we show the $\left\langle{\rm SFR}\right\rangle$
as a function of radio luminosity density for each of our
two redshift sub-samples. We calculate upper and lower limits for each 
luminosity density bin as we did for the whole sample. 
The upper and lower limits are indicated
by the grey shaded regions. Note due to the fact we only detect
4/15 sources in the moderate redshift sample we have to increase the bin size
by a factor of three compared to the high redshift sample. We also overlay the 
upper and lower limits for the whole 
of each sub-sample as indicated by the dashed lines. We see no evidence for any 
trend of mean SFR with radio luminosity for either sub-sample, although the 
constraints for the highest radio luminosity density bin of the high redshift 
sample are not so strong.

\section{Discussion}

We observe that radio-loud AGN in the distant Universe have an increasing 
mean SFR with cosmological look back time in the 
$25<\log(L_{1.4}/{\rm WHz}^{-1})<26.5$ radio luminosity density range. 
In the local Universe, $z<0.1$, the mean SFR of 
the 3CRR sample is $5-10$ times less than that in a moderate 
redshift sample, $0.4<z<0.9$. We note that the 3CRR sample was also selected
on FRII radio morphology which suggests we may not be comparing identical 
populations, and it may not be $100\,$per cent complete. 
Another recent study has examined the IR luminosities of bright radio sources
with {\it Herschel}-ATLAS observations of the GAMA-9h field. With a similar 
radio luminosity cut as our moderate redshift sub-sample, 
Hardcastle et al. (2010) find a mean SFR of between 
$20$ and $50\,{\rm M}_\odot$yr$^{-1}$, increasing across our moderate redshift 
bin (see Fig.~\ref{fig6}). 
\nocite{Hardcastle:10}
This range of mean SFRs is consistent with that found here, 
$18$$-$$41\,{\rm M}_\odot$yr$^{-1}$, allowing for the 
slightly different source selection, the different method of estimating 
IR luminosities and the fact these authors do not subtract any AGN 
contribution to the total IR luminosity.

We find the increase in mean SFR of radio-loud AGN hosts 
(parameterised as $\sim(1+z)^Q$, 
$Q=4.2\pm0.8$) to be greater than that of the IR luminosity function
which traces the evolution of the general star forming population. 
This greater rate of increase with redshift, compared to the regular
star forming population, suggests that some of the star formation may be
directly associated with the radio-loud AGN activity. The increase of mean SFR
with redshift of AGN is also seen in X-ray selected AGN 
\citep[e.g.][]{Lutz:10,Shao:10} and in a heterogenous sample of AGN 
\citep{Hatziminaoglou:10}. Alternatively, our results could reflect an 
increase in the stellar mass of the host galaxy, since high stellar mass galaxies
have SFRs which increase with redshift \citep[e.g.][]{Juneau:05}. This
interpretation would fit in with the recent \citet{Tadhunter:10} result
who find that at low redshifts, $z<0.7$, not all ULIRGs are massive enough 
to host radio-loud AGN. If the stellar masses of ULIRGs increase 
with redshift then ULIRGs would be more likely to host radio-loud AGN at 
higher redshifts.

While it is likely that the redshift information for the high redshift sample 
is biased toward sources that have bright $24\,\um$ flux densities
(see \S~\ref{sec:radio}), our approach of determining a range of mean SFRs
given two extreme assumptions alleviates much of the concern about selection bias.
The remaining principle source of uncertainty is the S-cubed model, used
to quantify how complete our sub-samples were. As discussed earlier, our 
uncertainties in S-cubed are very conservative.
S-cubed treats the AGN and SFGs as separate populations, i.e. it does not 
include hybrid radio sources exhibiting both processes simultaneously.
We can thereby compare the expected number of radio-loud AGN 
regardless of whether there is ongoing star formation in their hosts or not.
The contribution to the SFR density of the host galaxies of radio-loud AGN in 
the high redshift bin is interesting as the SFR density at this epoch is 
dominated by LIRGs and ULIRGs \citep{LeFloch:05, Seymour:10}. 
As $0.1-0.5\,$per cent of the SFR density consists of LIRGs and ULIRGs which 
host the
radio-loud AGN, we can infer a duty cycle of $0.001-0.005$ for radio-loud AGN 
activity in such sources, assuming that each LIRG and ULIRG goes through at 
least one radio-loud phase.
The typical time-scale of a radio-loud phase of an AGN is around $\sim10\,$Myr 
for extended radio sources \citep{Miley:80} and likely shorter for the less 
luminous sources with smaller radio lobes
considered here. Given this lifetime and the estimated duty cycle of $0.001-0.005$
we can estimate that LIRGs and ULIRGs undergo a radio-loud AGN phase every 
$2-10\,$Gyr. Hence, during the $3\,$Gyr time 
span covered by the high redshift sub-sample we could expect perhaps
one major phase of radio-loud AGN activity at a rate similar to that expected 
from major mergers \citep{Hopkins:10}. 

The feedback models which quench star formation by evoking a radio-loud phase 
\citep[e.g.][]{Croton:06,Bower:06} are most important at late times, 
i.e. below $z<1$, but they must occur at higher redshifts in order to
prevent the most massive galaxies, formed at early times, from growing 
significantly more. However, in this work we observe many AGN in our high 
redshift sub-sample which are in a state equivalent to the `radio-mode' 
feedback of \citet{Croton:06,Bower:06} and simultaneously
have very high SFRs while feedback processes are predicted to be 
occuring.

\section{Conclusions}

We have examined the incidence of  far-IR emission and inferred SFR of 
luminous radio-loud AGN in a moderate redshift, $0.4<z<0.9$, and a high 
redshift sub-sample, $1.2<z<3$, as well as a local, $z<0.1$, comparison 
sample. We have:

\begin{itemize} 
\item constrained the mean SFR of radio-loud AGN to be $3.4$$-$$4.2$, $18$$-$$41$ 
and $80$$-$$581\,{\rm M}_\odot$yr$^{-1}$ for the local, moderate and high 
redshift samples respectively, hence, we measure the evolution of the 
mean SFR to be $\sim(1+z)^{4.2\pm0.8}$,
\item observed no strong trends of SFR with radio luminosity in any redshift bin,
\item estimated that the host galaxies of radio-loud AGN in the high redshift 
sub-sample contribute $0.1-0.5\,$per cent to the total SFR density at that epoch 
and if all LIRGs and ULIRGs have a radio-loud phase we infer a duty cycle of 
$0.001-0.005$ in such sources.
\end{itemize}

These results  demonstrate that in the distant Universe 
a considerable amount of star formation is occuring in galaxies hosting a 
radio-loud AGN, consistent with the frequent evidence for high SFRs 
in classic high redshift radio galaxies.
The mean SFR evolves more quickly than the IR luminosity function
implying that some of the star formation is directly related to the radio-loud 
AGN activity.
Both starburst and active nuclear processes have 
relatively short time-scales so their co-existence in many objects suggests that 
bursts of star formation and jet activity either are quite common or connected
via `feedback'. But is the jet 
initiating or quenching star formation, or are the processes independent? We 
cannot answer such questions here, but we shall be able to do so 
with follow-up of individual sources (to search for outflows of jet-triggered 
star formation or for mergers triggering both) and with the 
huge sample that will be provided by the full HerMES data set combined with 
improved redshift information.

\section*{Acknowledgments}
We thank the referee for their useful comments which improved the presentation
of this paper.
NS thanks Carlos De Breuck, Martin Hardcastle, Curtis Saxton and Clive 
Tadhunter for useful discussions.
SPIRE has been developed by a consortium of institutes led by Cardiff
Univ. (UK) and including Univ. Lethbridge (Canada); NAOC (China); CEA, LAM
(France); IFSI, Univ. Padua (Italy); IAC (Spain); Stockholm Observatory
(Sweden); Imperial College London, RAL, UCL-MSSL, UKATC, Univ. Sussex
(UK); Caltech, JPL, NHSC, Univ. Colorado (USA). This development has been
supported by national funding agencies: CSA (Canada); NAOC (China); CEA,
CNES, CNRS (France); ASI (Italy); MCINN (Spain); SNSB (Sweden); STFC (UK);
and NASA (USA). The data presented in this paper
paper will be released through the Herschel Database in Marseille HeDaM 
(hedam.oamp.fr/HerMES). 


\bsp

\label{lastpage}

\end{document}